# Time and Frequency Resolved Nanoscale Chemical Imaging of 4,4′-Dimercaptostilbene on Silver

*Patrick Z. El-Khoury,* Dehong Hu, and Wayne P. Hess**

Physical Sciences Division, Pacific Northwest National Laboratory, P.O. Box 999, Richland, WA 99352, USA

## ABSTRACT

Non-resonant tip-enhanced Raman images of 4,4′-dimercaptostilbene on silver reveal that different vibrational resonances of the reporter are selectively enhanced at different sites on the metal substrate. Sequentially recorded images track molecular diffusion within the diffraction-limited laser spot which illuminates the substrate. In effect, the recorded time resolved ($\Delta t = 10$ s) pixelated images (25 nm x 8 $cm^{-1}$) broadcast molecule-local field interactions which take place on much finer scales.

## KEYWORDS

Raman Scattering, SERS, TERS, Nanoscale, Chemical Imaging, Local Field, Dimercaptostilbene

The ability to harness the resonant interaction between electromagnetic fields and plasmonic Eigenmodes in metal nanostructures has enabled several emerging technologies. The localization of surface plasmons using engineered metal nanostructures and their interaction with molecular polarizability tensors have afforded single molecule detection sensitivity[1,2,3] in surface-enhanced Raman scattering (SERS),[4,5,6] and more recently, chemical imaging within one molecule[7] through tip-enhanced Raman scattering (TERS).[8,9] Single molecule detection sensitivity can be achieved by taking advantage of electromagnetic enhancement factors exceeding $10^{10}$, amenable at plasmonic nanojunctions where molecules are sited and interrogated. Herein, we closely examine the information content in time ($\Delta t$ = 10 s) and frequency (8 $cm^{-1}$ resolution) resolved TERS images (25 nm spatial resolution) obtained by repeatedly scanning a 25 nm gold AFM probe through a diffraction-limited laser spot illuminating a corrugated silver substrate coated with 4,4′-dimercaptostilbene (DMS).

Several challenges are faced in attempting to faithfully account for the enhanced inelastic photon scattering response at plasmonic nanojunctions. They mainly stem from our inability to control structures and constructs over length scales on the order of/less than a nanometer. The most mundane of considerations has to do with the inhomogeneity of electric fields around nanometeric asperities sustained on metal surfaces and tips.[10,11] In principle, even when ideal nanojunctions (e.g. the prototypical nanosphere dimer) are considered, the enhanced local fields were found to vary over typical molecular length scales.[10] This has been associated with the observation of surface-enhanced Raman optical activity at nanojunctions, which necessitates the expansion of molecular polarizability to account for the recorded surface-enhanced electric dipole-, electric quadrupole-, and magnetic dipole-coupled Raman spectra.[11] Another complication arises from the nanoparticle separation distance-dependent plasmonic response.

Namely, when the separation distance is decreased such that conductive overlap is established between the nanoparticles, charge transfer plasmon (CTP) modes compete with the more familiar dipolar bonding dimer plasmon (BDP) modes.[12,13] SERS from a single biphenyl-4,4'-dithiol (BPDT) molecule linking two silver nanospheres uncovered the transition from BDP to CTP modes throughout the fusion of the metal nanojunction.[11] The transition to the realm of current-carrying plasmons was marked by a dramatic increase in scattering activity, accompanied by significant spectral broadening. Through TERS trajectories recorded at a nanojunction formed by a gold AFM tip in contact with a silver surface coated with the same dithiol, we recently demonstrated[14] how the aforementioned spectral signatures can be used to monitor the breaking and making of chemical bonds with a method that affords single bond detection sensitivity.

Beyond conceptual and practical complications stemming from the nonlocal plasmonic response which varies over molecular length scales, time-dependent spectral variations in sequences of Raman spectra recorded at plasmonic junctions comprise yet another degree of complexity. Recent works have illustrated that when the optical response is nascent from a single scatterer (or a few), the full tensor nature of Raman scattering comes into play.[7,11,15,16] Namely, the relative intensities in single molecule non-resonant Raman spectra were correlated with molecular orientation relative to the enhanced local electric field; in the limit where molecular reorientation does not affect the spectroscopic properties of the system as a result of specific molecule-metal interactions.[17] In this study, we demonstrate how such effects govern TERS micro-spectroscopy. Frequency resolved TERS images of a corrugated silver substrate coated with DMS reveal that different vibrational resonances are selectively enhanced at different locations of the substrate. In effect, the pixelated images (25 nm x 8 $cm^{-1}$) reconstructed at different vibrational resonances of the reporter broadcast intimate details about molecule-enhanced local electric field interactions taking place on much finer scales; which are otherwise only accessible in experiments conducted under ultrahigh vacuum and ultralow temperatures.[7,18,19] Moreover, sequentially recorded TERS images allow us to directly visualize molecular diffusion between various hotspots localized within the laser spot.

Following laser-tip alignment, the AFM probe is scanned back and forth (± 1.25 μm in 25 nm increments) through the ~250 nm laser spot laser spot illuminating the DMS-coated silver substrate. Raman spectra were recorded at every tip position, with an integration time of 0.1s per spectrum. Figure 1A is a contour plot representation of 10 Raman images recorded from 10 consecutive passes of the AFM probe through the laser spot. The time and tip travel distance-dependent scattering signals at select vibrational resonances of the reporter are plotted in Figure 1B. When the tip resides outside the laser spot, only surface-enhancement is operative. In this case, it is generally accepted that the overall response is dominated by Raman scattering from molecules residing at various hotspots within the illuminated region.[20,21] This accounts for the constant baseline SERS signal observed at the different resonances. As the tip approaches the laser spot, the molecular signatures are further enhanced. The additional TERS signal arises from a smaller number of scatterers residing at hotspots sustained within the effective near-field scattering area, now further enhanced by the metal probe.[22] The TERS/SERS contrast can be readily discerned by inspecting Figures 1C-1D. The constant background signal in Figure 1C corresponds to the average SERS scattering signal, whereas the enhanced signals at 0.975 ± 0.1 μm are a result of further tip enhancement. In the ensuing discussion, we illustrate how these signals can be used to locate, image, and characterize hotspots within the illuminated region.

Several intricate considerations arise in trying to infer the vector components of the incident and scattered radiation fields at the non-stationary tip-sample nanojunction.[16,23] Whereas polarization in the z direction of light propagation (along the tip axis) is required to affect TERS, the polarization of the driving field in our inverted microscope setup is transverse to the tip axis (in the xy plane of the substrate). In principle, this does not allow the incident polarization to efficiently couple into the metal probe, and tip-enhancement is not possible. Nevertheless, the incident beam is focused onto the sample surface using a high numerical aperture objective (1.3 NA), which partially converts transverse fields to polarization parallel to the tip axis. In an idealized scenario, the corresponding TERS images would effectively trace the spatial distribution of the z component of the electric field. This is not what we observe. In Figure 2, we zoom into the TERS signals in the 0.975 ± 0.1 μm region corresponding to the first pass of the AFM probe through the diffraction-limited laser spot, and display the top 10% brightest positions to visualize the location of hotspots on the silver substrate. The corresponding Raman images are expanded at frequency shifts corresponding to different vibrational resonances, a vector representation of which is shown in the insets, see Figures 2A-D. Note that Figure 2B is a zoomed in version of Figure 1C (in all 3 dimensions of space, frequency, and scattering activity) reconstructed using the brightest Raman-active vibrational normal mode of DMS assigned to its aromatic C=C stretching vibration coupled to in-plane vibrations of the aromatic CHs. When the ~1580 $cm^{-1}$ mode is used as a probe of TERS activity, the most intense signals are located at 175 and 200 nm . When other vibrational normal modes of DMS are used as probes of TERS activity, it can be immediately recognized that optimal TERS enhancement occurs at different sites (hotspots) within the illuminated region of the substrate. This is also evident in Figures 3, discussed below. These results can be rationalized on the basis of our recent reports. In the absence of specific molecule-metal interactions[17] which perturb the electronic structure of DMS, and in the dipole coupling limit,[11] the scattering tensor which controls TERS intensity can be written as[16]

$$S_n^2 = \sum_n \left| E_s^L \alpha_n'(\Omega) E_i^L \right|^2$$

where $E_{i,s}^L$ are the enhanced incident and scattered local radiation fields, $\alpha_n'$ is the molecular polarizability derivative tensor for the $n^{th}$ vibrational Eigenstate, and $\Omega = \{\alpha, \beta, \gamma\}$ are the Euler angles which determine molecular orientation relative to the local fields. In this framework, the intensity of the $n^{th}$ vibrational state is dictated by the relative orientation of a single or a few molecules with respect to vector components of $E_{i,s}^L$, defined by the interaction of the tip with different regions of the corrugated metal surface. In principle, the same analysis carried out in ref [16] can be employed herein to infer the relative orientation of scatterer(s) with respect to the local radiation fields at the different tip positions. That said, we refrain from duplicating the prior analyses to which the reader is referred,[11,15,16] and restrict the present report to the imaging facet of our results. In this regard, notice how the predominant elements of the molecular polarizability derivative tensor of the brightest 1580 cm$^{-1}$ mode effectively limit the overall spatial resolution and contrast in the TERS image shown in Figure 2B when compared to its analogues shown in Figures 2A,C, and D. Namely, irrespective of the complexity of the interstitial enhanced local fields, there are always projections of the tensor elements of this mode onto the fields which effectively lead to its selective enhancement. Combined with our prior analysis,[15] this result suggests that normal modes featuring comparable magnitudes of their spherical and/or anisotropic polarizability tensor elements would result in better TERS/SERS imaging contrasts.

The second and third passes of the AFM tip over the laser spot can also be visualized in perspective of various vibrational modes of DMS in Figures 3. These images are sequentially recorded 10 (Figure 3B) and 20 (Figure 3C) seconds after the first scan (Figures 2 and 3A) of the same region is performed. We notice that different hotspots light up in the second image, at the expense of diminished/absent TERS activity at the original positions observed from the first pass of the metal probe over the illuminated region of the substrate. This is followed by signal recovery at the initial hotspots in the third image along the sequence, see Figure 3. The second and third images reveal additional, previously unpopulated (optically silent) hotspots. We attribute these observations to molecular diffusion within the laser spot. It's worth noting that physical contact of the molecules with the tip is excluded on the basis of time sequences of contact mode tip-enhanced Raman point spectra (not shown), which yielded time-dependent scattering signals reminiscent of CTP-enhanced TERS previously observed from a structurally similar dithiol.[14] That said, the observed process can be driven by a collation of effects ranging from optical forces at the tip-sample nanojunction through electrostatic interactions between the molecules and the transient tip. Although the exact driving force(s) for diffusion cannot be ascertained at present, our observations of (i) reversible population of the same hotspots at the four different frequencies, and (ii) the population of new hotspots within the probed region both suggest that molecular diffusion is taking place. In principle, this assignment also explains the observed variations in optimal TERS activity towards the center of the laser spot (see Figure 1B), all throughout (i) the sequence consisting of 10 consecutively collected TERS images, and (ii) tens of similar sequences collected from various regions of the substrate. We will address this issue in greater detail in follow-up works which will also expand upon the various aspects of TERS micro-spectroscopy touched in this present letter.

## METHODS

The samples were prepared using a 0.1 mM solution of 4,4'-dimercaptostilbene (DMS, Sigma-Aldrich) in ethanol (Gold Shield, 200 proof), spin-casted onto ~15 nm Ag films evaporated on a 0.1 mm-thick microscope slide by arc-discharge physical vapor deposition (target: Ted Pella Inc., 99.99% purity). The typical surface roughness measured on freshly evaporated silver films indicates a root-mean-square height distribution of ~3 nm. This was followed by rigorously washing the substrate with ethanol, and vacuum evaporation to rid the sample of residual solvent. TERS measurements were conducted under ambient laboratory conditions using an atomic force microscope (Nanoscope IIIa, Veeco Metrology) operating in non-contact mode, mounted on an inverted optical microscope (Axiovert 200, Zeiss). The incident 514 nm continuous wave laser (Innova 300, Coherent) is attenuated to 45 $\mu W/\mu m^2$ using a variable neutral density filter wheel, reflected off a dichroic beamsplitter, and focused onto the sample using an oil-immersion objective (1.3 NA, 100 X). The gold AFM probe (25 nm cone radius) was aligned with the diffraction-limited laser spot by maximizing the nascent molecular signal at ~1580 $cm^{-1}$. The scattered radiation is collected through the same objective, transmitted through a beamsplitter, and filtered through a long pass filter. The resulting light is detected by a liquid nitrogen cooled charge coupled device coupled to a spectrometer (Holespec f/1.8i, Kaiser Optical System). The effective spectral resolution of our instrument is ~8 $cm^{-1}$.

## FIGURES

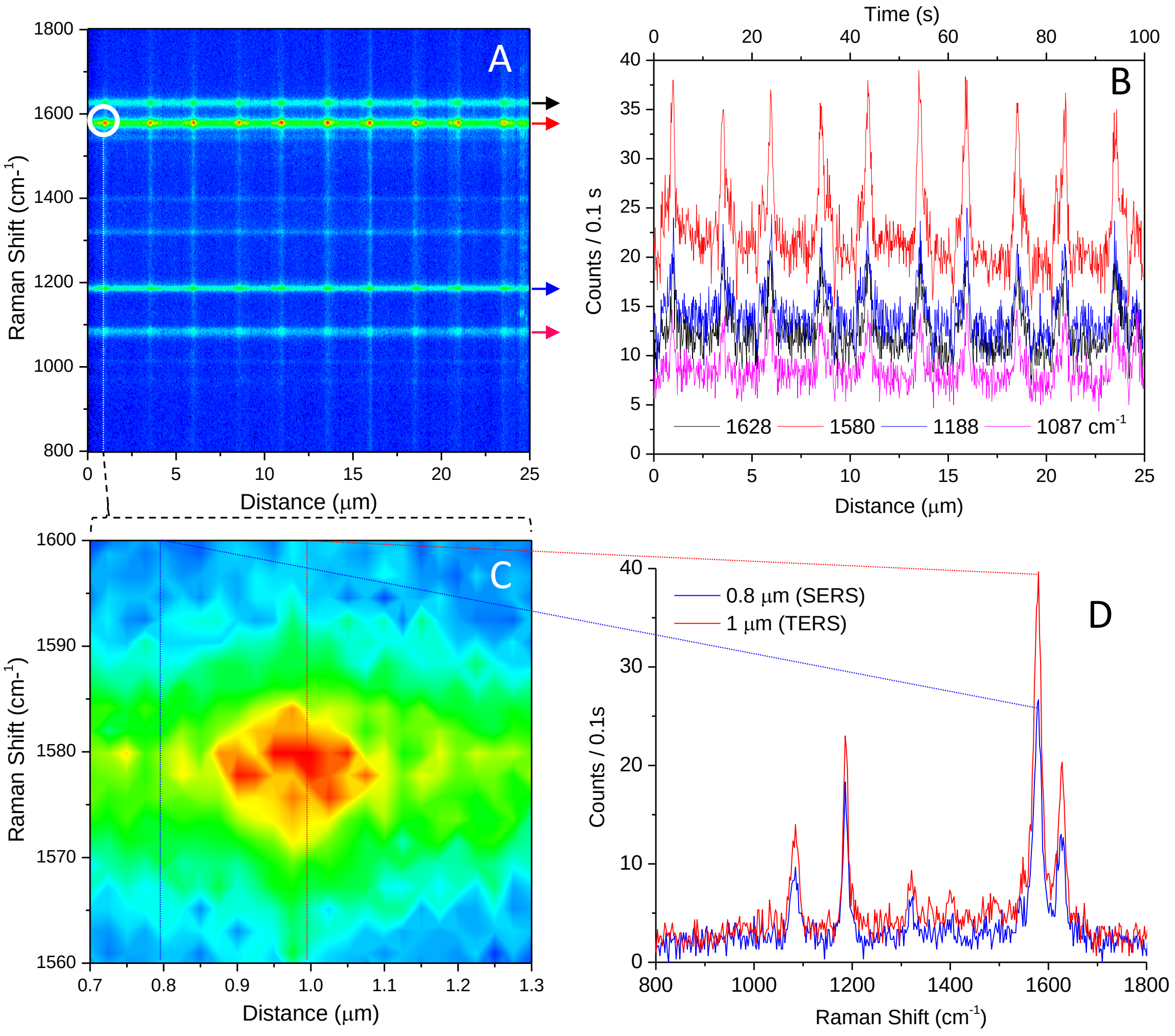

**Figure 1.** A) AFM tip travel-distance dependent SERS and TERS spectra recorded by scanning a gold AFM tip ± 1.25 µm (in 25 nm increments) from the center of a diffraction limited laser spot illuminating a corrugated silver substrate coated with DMS. B) The corresponding evolution of scattering activity at select vibrational resonances of DMS. C) Expansion of the region highlighted in A). D) SERS/TERS spectra in the absence/presence of the tip in the effective probing area defined by the laser spot.

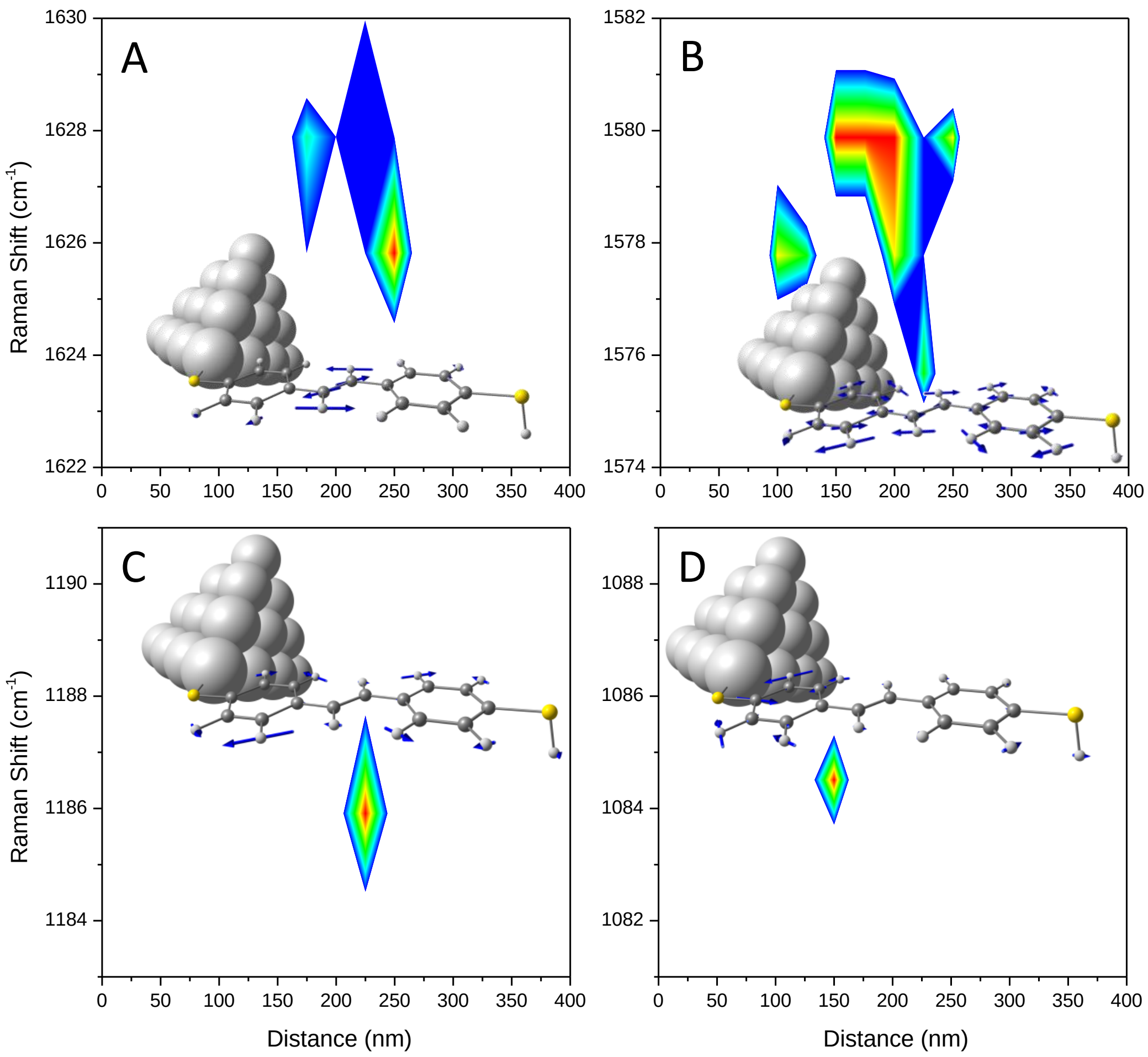

**Figure 2.** TERS imaging in perspective of the different vibrational modes of DMS, obtained from the first pass of the AFM probe through the laser spot. A vector representation of the normal modes is shown in the insets.

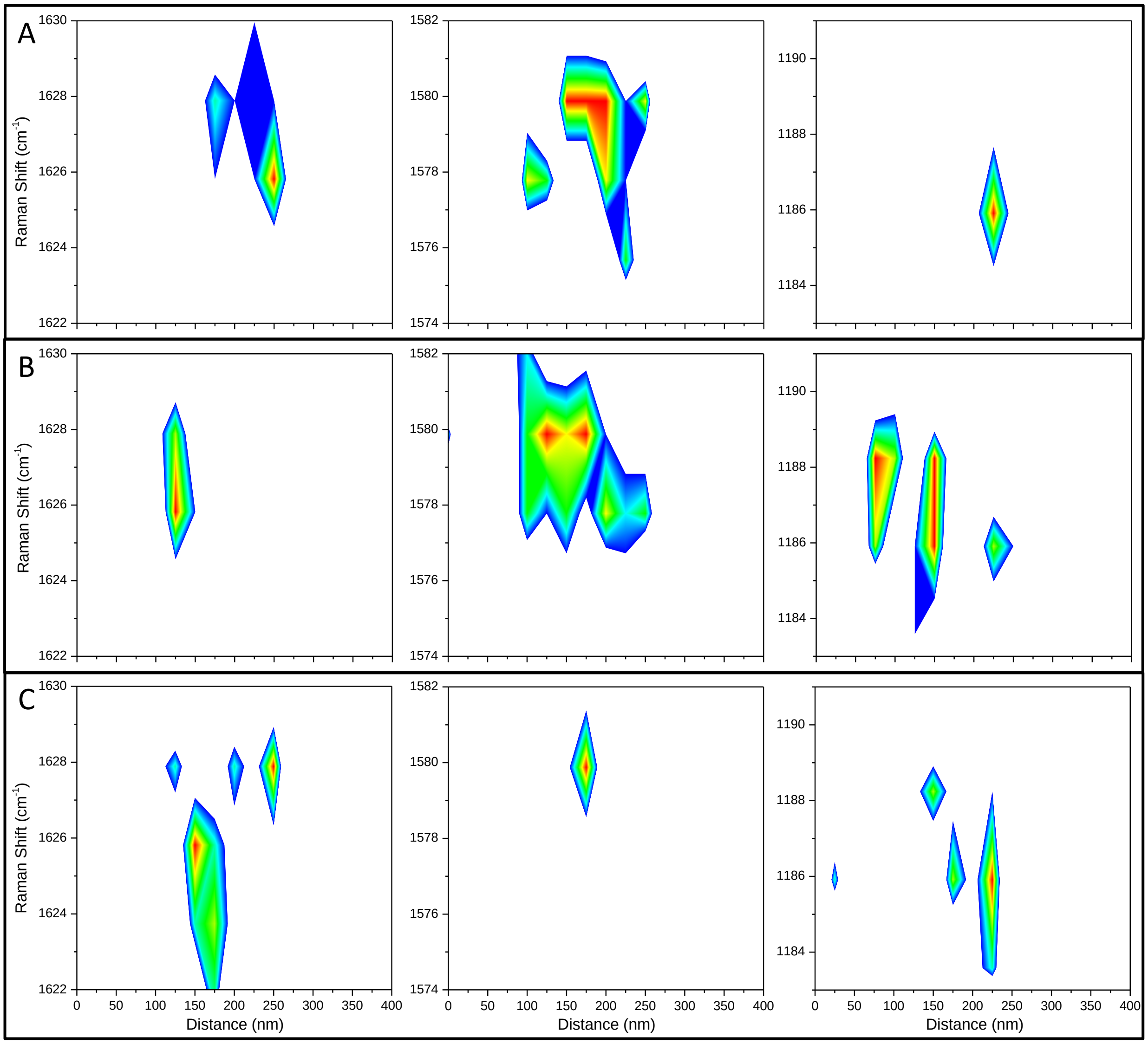

**Figure 3.** TERS imaging in perspective of different vibrational modes of DMS, obtained from the first (A), second (B), and third (C) passes of the AFM probe through the laser spot.

## AUTHOR INFORMATION

**Corresponding Authors**

*patrick.elkhoury@pnnl.gov (PZE), wayne.hess@pnnl.gov (WPH)

## ACKNOWLEDGMENTS

WPH acknowledges support from the US Department of Energy (DOE), Office of Basic Energy Sciences, Division of Chemical Sciences, Geosciences & Biosciences. PZE acknowledges support from the Laboratory Directed Research and Development Program through a Linus Pauling Fellowship at Pacific Northwest National Laboratory (PNNL) and an allocation of computing time from the National Science Foundation (TG-CHE130003). This work was performed using EMSL, a national scientific user facility sponsored by DOE's Office of Biological and Environmental Research and located at PNNL. PNNL is a multiprogram national laboratory operated for DOE by Battelle.